%% file: visco_active_PRE_v1.tex
\documentclass[pre]{revtex4}
\include{tex_options}

\newcommand{\symmtraceless}[1]{\vbox{\ialign{##\crcr
       \vrule height0.4pt depth2pt
       \hrulefill
       \vrule height0.4pt depth2pt
       \crcr\noalign{\kern1pt\nointerlineskip}
       $\hfil\displaystyle{#1}\hfil$\crcr}}}
\newcommand{\stl}[1]{\symmtraceless{#1}}

\begin{document}
%%%%%%%%%%%%%%%%%%%%%%%%%%%%%%%%%
\title{Non-linear rheology of active particle suspensions: Insights from an analytical approach}
\author{Sebastian Heidenreich}
\email{s.heidenreich1@physics.ox.ac.uk}
\affiliation{Rudolf Peierls Centre for Theoretical Physics, University of Oxford, 1 Keble Road, Oxford OX1 3NP, United Kingdom}
\affiliation{Institut f\"ur Theoretische Physik, Sekr. EW7-1, Technische Universit\"at 10623 Berlin, Hardenbergstrasse 39, Berlin, Germany}
\author{Siegfried Hess}
\affiliation{Institut f\"ur Theoretische Physik, Sekr. EW7-1,  Technische Universit\"at 10623 Berlin, Hardenbergstrasse 39, Berlin, Germany}
\author{Sabine H. L. Klapp}
\affiliation{Institut f\"ur Theoretische Physik, Sekr. EW7-1, Technische Universit\"at 10623 Berlin, Hardenbergstrasse 39, Berlin, Germany}
\date{\today}
\begin{abstract}
We consider active suspensions in the isotropic phase subjected to a shear flow. Using a set of extended hydrodynamic equations we derive a variety of {\em analytical} expressions for rheological quantities such as shear viscosity and normal stress differences. In agreement to full-blown numerical calculations and experiments we find a shear thickening or -thinning behaviour depending on whether the particles are contractile or extensile. Moreover, our analytical approach predicts that the normal stress differences can change their sign in contrast to passive suspensions.
 \end{abstract}

\pacs{
4
}
\maketitle
%%%%%%%%%%%%%%%%%%%%%%%%%%%%%%%%%

\section{Introduction}
\label{intro}
Within the last years the physics of active materials has become a focus of growing interest. In an active system, each particle (constituent) "constantly absorbs
energy from its surroundings or from  an internal fuel tank (such as adenosinetriphosphate) and dissipates it in a process of carrying out translational or rotational motion" \cite{Hatwalne}. The resulting, continuous
energy flux drives an active system out of equilibrium even in a steady state. This is in sharp contrast to the physics of ''passive'' soft-matter systems where non-equilibrium behavior typically occurs only in the presence of an {\em external} driving force (such as shear).

Active systems often occur in biological contexts, examples being acto-myosin filaments, cytoskeletal gels interacting with motors, bacterial colonies, and swarming fishes \cite{Ramaswamy, Kruse, Liverpool}. Moreover, there are also non-biological representatives such as
vibrated granular rods \cite{Ramaswamy_Science, Tsimring} or remotely powered, self-propelled particles \cite{Velev}.

The individual motion of an active particle through the suspension gives rise to flow fields which, on long time scales, can be modeled by force dipoles $\mathbf{n}$ \cite{Winet,Hatwalne}.  The combination of this directed motion and interactions between active particles (e.g., excluded volume interactions or hydrodynamic interactions) generates rich collective behaviour with some similarities to that of ordinary
''passive'' liquid crystals. However, active systems (''living liquid crystals'' \cite{Gruler}) have fascinating additional features such as an instability of homogeneous, nematic states \cite{Simha}
or the occurrence of {\em spontaneous} flow states in films above a critical thickness \cite{Porst,Joanny,Rao,YeomansMarenduzzo,Yeomans1,Giomi}.

Modeling these phenomena on a {\em microscopic level} is still a challenge \cite{Liverpool}. As an alternative, the collective  behavior is often studied  via a coarse-grained continuum approach familiar from liquid-crystal theory.
Within this framework the relevant dynamic variables are the tensorial orientational order parameter $Q_{\mu \nu}=\langle \stl{n_\mu n_\nu}\rangle$
\cite{YeomansCates, Hatwalne} (where $n_\mu$ describes the particle orientation and
$\stl{\ldots}$  indicates the symmetric traceless part of a tensor),  the velocity field, the stress tensor $\sigma_{\mu \nu} $, the density (if the latter is allowed to vary), and optionally, the polarization of the system \cite{Giomi,Marchetti1,Marchetti2,Hatwalne,Ramaswamy,YeomansMarenduzzo}. The latter becomes particularly important in systems of so-called ''movers''
such as fishes and bacteria. The resulting hydrodynamic equations are then supplemented by additional terms which take into account
the active nature of the system (see, e.g., Refs.~\cite{Joanny,YeomansCates,Hatwalne}). As a minimal {\em ansatz} it is often assumed that the individual force dipoles generate
a contribution to the stress tensor which is linear in $Q_{\mu \nu}$. The prefactor can be either positive or negative describing, respectively, ''contractile'' systems (such as suspensions of {\em alga Chlamydomonas Reinhardii}, cytoskeletal rods, actomyosin)
or ''extensile'' particles (e.g., {\em E.~Coli} bacteria).

In the present work we use a hydrodynamic approach to explore the non-linear response of an active material to an {\em external} shear flow.  We focus on a non-polar
material composed of ''shakers'' (such as melanocytes or symmetric rods) and neglect any variations in the number density. The passive contributions to the equations of motion are modeled in the spirit of our earlier work on passive liquid-crystalline systems (see, e.g., \cite{Borgmeyer_Hess,Heidenreich_diss,Heidenreich_dip}). Activity is then incorporated by the above-mentioned stress term plus the lowest-order contribution
to the order parameter itself. Previous theoretical investigations on the rheology
of active liquid-crystals have already indicated that their flow behavior is more complex than that of the corresponding passive system.
On top of the above-mentioned spontaneous flow  one observes, e.g., activity-induced shear band formation \cite{Giomi,YeomansCates}, hydrodynamic instabilities \cite{Gruler,Simha}, fluid mixing \cite{Shelley}, pattern formation
in steady states \cite{Tsimring1,Zimmermann} and cell-sorting phenomena \cite{Chate}.

Contrary to many of these studies, which often involve quite complex numerical investigations, we aim to derive {\em analytical} results. The focus is on experimental accessible quantities such as the shear viscosity and the normal stress differences.
Our motivation to approach these quantities from an analytical side is to establish simple ''rules'' for the role of, e.g., the prefactor of the activity-induced pressure term, on the
non-Newtonian viscosity. We hope that the physical insight gained by such simple rules may be particularly interesting for the interpretation of future experiments. Moreover, the results can also serve
as benchmark for more complex numerical studies.
Of course, the derivation of explicit expressions is impossible for the full, nonlinear set of hydrodynamic equations. The main
assumption in our work is that the system is close to the passive isotropic state.
 From a physical point of view this means that we are essentially restricted to the investigations of states which are
isotropic in passive equilibrium in the absence of external flow. Despite this assumption, our theory shows a variety of non-linear flow phenomena such as
shear thickening and shear thinning in dependence of the activity parameter. Moreover, the theory
yields novel results for the normal stress differences, which are a characteristic rheological feature of any soft-matter system beyond the linear-response (Newtonian)
regime.

The remainder of this paper is organized as follows. In section \ref{model} we present our set of hydrodynamic equations and motivate (following the arguments in
earlier studies) the extra terms due to activity. Section \ref{secviscosity} deals with the shear viscosity (in a slab geometry),
focussing on a spatially homogeneous system. Here we present analytical expressions and discuss our results in the light of earlier, fully numerical studies. Section~\ref{stress} is devoted to the normal stress differences on the same level of approximation.
\section{Model Equations}
\label{model}
In the continuum limit the orientational dynamics of active particle suspension is governed by a coarse-grained equation of motion
for the second-rank order parameter $Q_{\mu \nu}$ \cite{Hatwalne}.
The starting point are the well-established relaxation equations
for passive liquid crystals \cite{Heidenreich_dip}, which may then be supplemented by terms reflecting the activity. Including these effects to lowest order \cite{Hatwalne, Simha},
the resulting equation for the time derivative of $Q_{\mu \nu}$ is given by
\begin{eqnarray}
\label{modelEq}
(\partial_t+ v_\lambda \nabla_\lambda  ) Q_{\mu \nu}&=& \stl{ \epsilon_{\mu \lambda \kappa} \omega_\lambda Q_{\kappa \nu}}+
2 \kappa \stl{\Gamma_{\mu \lambda} Q_{\lambda \nu}}  -\sqrt{2} \tau_\mathrm{Q}^{-1}  \tau_\mathrm{Qp}  \Gamma_{\mu \nu}\nonumber \\
&&-\tau_\mathrm{Q}^{-1} \Phi_{\mu \nu} +\frac{\xi_0^2}{\tau_Q} \triangle Q_{\mu\nu}+\lambda Q_{\mu\nu},
\end{eqnarray}
where we use the Einstein summation convention and the notation $v_\mu$ for the velocity. The first three terms on the right-side of Eq.~(\ref{modelEq}) describe the impact of the flow on the
order parameter.
Specifically, $\Gamma_{\mu \nu}=\stl{\nabla_\mu v_\nu}$ is the strain rate, $\omega_{\mu \nu}$,the vorticity, and $\tau_\mathrm{Qp}$ and $\tau_\mathrm{Q} $ are
relaxation time coefficients, respectively. In the subsequent term in Eq.~(\ref{modelEq}), $\Phi_{\mu \nu}=\delta \Phi/\delta Q_{\mu \nu} $ is the ''molecular field'',  which is described by the
derivative of the Landau-de Gennes (LG) free energy $\Phi$ of a homogeneous, liquid-crystalline system,
\begin{equation}
 \Phi=\frac{1}{2}A(T)Q_{\mu \nu}Q_{\mu \nu}-\sqrt{\frac{2}{3}}B Q_{\mu \lambda}Q_{\lambda \nu}Q_{\mu \nu}+\frac{1}{4}C (Q_{\mu \nu}Q_{\mu \nu})^2,
\end{equation}
%*
where $A(T)=\alpha(T-T_\mathrm{c})$ (with $T$ being the temperature) changes its sign at the isotropic-nematic spinodal temperature, $T_\mathrm{c}$.
Note that the definition of a free energy of an active system is, strictly speaking, impossible due to its non-equilibrium character. We thus consider $\Phi$ as the system's
free energy in the ''passive phase''. The next (gradient) term in Eq.~(\ref{modelEq})
is again borrowed from equilibrium physics; it results from the increase of free energy due to distortions in the one-constant approximation. The prefactor of this ''elastic''
contribution corresponds to the squared elasticity length, $\xi_0^2$. So far, the terms in Eq.~(\ref{modelEq})
correspond to those for passive systems (see, e.g., \cite{Heidenreich_dip}).
It is the last term in Eq.~(\ref{modelEq}) which corresponds
to an {\em ansatz} for the effect of
activity on the orientational order, with $\lambda$ being a coupling parameter. For $\lambda>0$ ($\lambda<0$) the order parameter $Q_{\mu \nu}$ increases
(decreases) in time; the case $\lambda>0$ is therefore sometimes referred to as ''self-aligning''. For a microscopic interpretation of the coefficient $\lambda$, see \cite{Marchetti6}.

Some further interpretation arises when we consider this term, which is linear
in $Q_{\mu\nu}$, together with the linear term resulting from the derivative of the squared term in the LG free energy. This yields a contribution
$\left(-\tau_\mathrm{Q}^{-1}A(T)+\lambda\right)Q_{\mu\nu}=-{\tau_\mathrm{Q}}^{-1}\delta \tilde{\Phi}/\delta Q_{\mu \nu}$, where
$\tilde{\Phi}$ is a pseudo free energy defined by
\begin{equation}
\label{linear_term}
\tilde\Phi=\frac{1}{2}\left(A(T)-\lambda\tau_\mathrm{Q}\right)Q_{\mu \nu}Q_{\mu \nu}.
\end{equation}
From this consideration we see that $\lambda$ influences the onset of nematic ordering upon cooling the system from high temperature (i.e., large values of $A$).
Specifically, a positive value of $\lambda$ effectively {\em increases} the spinodal
temperature and thus stabilizes nematic ordering with respect to the passive case.
On the other hand, $\lambda<0$ stabilizes the isotropic (i.e., orientationally disordered) state \cite{YeomansMarenduzzo}. We also see from Eq.~(\ref{linear_term}) that
for $\lambda>\tau_\mathrm{Q}^{-1}A(T)$ the isotropic state is unstable (i.e., $\delta^2\tilde{\Phi}/(\delta Q)^2<0$). Finally, Eq.~(\ref{linear_term}) shows that
the order of magnitude of $\lambda$ is given by the inverse of the relaxation time $\tau_\mathrm{Q}$.

The fluid velocity field obeys the generalized Navier-Stokes equations
\begin{equation}
\label{velocity}
 \rho(\partial_t +v_\lambda \nabla_\lambda)v_\mu=2\eta_{iso} \partial_\lambda \Gamma_{\lambda \mu}+\partial_\lambda \sigma_{\lambda \mu},
\end{equation}
where $\rho$ is the fluid density, and $\eta_\mathrm{iso}$ is the first Newtonian viscosity of the passive system.
Following earlier studies (see, e.g. \cite{Simha,Yeomans1}) we model the stress tensor $\sigma_{\lambda \mu}$ as a sum of a passive part, $\sigma^\mathrm{p}$, and an active part, $\sigma^\mathrm{a}$, i.e.,
\begin{equation}
\label{stress_sum}
\stl{\sigma}_{\mu \nu}=\stl{\sigma}^\mathrm{p}_{\mu \nu}+\stl{\sigma}^\mathrm{a}_{\mu \nu}.
\end{equation}
The passive part is given by the corresponding liquid-crystal expression \cite{Borgmeyer_Hess}
\begin{eqnarray}
\label{stressEq}
 \stl{\sigma}^\mathrm{p}_{\mu \nu}&=&
-\frac{\rho}{m}k_\mathrm{B}T\sqrt{2} \left( \frac{\tau_\mathrm{Qp}}{\tau_\mathrm{Q}} (\Phi_{\mu \nu}-\triangle Q_{\mu\nu}) \right. \\
&&- \left. \sqrt{2} \kappa \stl{ Q_{\mu \lambda} \Phi_{\lambda \nu}} +\sqrt{2}\kappa \, \xi_0^2 \, \stl{Q_{\mu \lambda}\triangle Q_{\lambda \nu}} \right),
\end{eqnarray}
which can be derived from the free energy of the system and the principles of irreversible thermodynamics. The presence of the active term,
$\sigma^\mathrm{a}$,
results from the fact that the self-propelled particles in an active system induce flow fields which are dipolar in character \cite{Hatwalne, Simha}. Here we use the simplest ansatz (for higher-order corrections see \cite{Marchetti3})
\begin{equation}
\label{stress_active}
\stl{\sigma}^\mathrm{a}_{\mu \nu}=-\frac{\rho}{m}k_\mathrm{B}T\zeta Q_{\mu \nu}.
\end{equation}
The proportionality constant distinguishes between extensile ($\zeta >0$) and contractile ($\zeta < 0$) active flow behavior \cite{ Hatwalne,Yeomans1,YeomansCates} and is characterized by the strength of the elementary force dipoles. From Eqs.~(\ref{stress_active}) and (\ref{velocity}) we then see that an extensile system {\em opposes} the
external flow (described by $\Gamma_{\nu\mu}$), whereas a contractile system {\em enhances} it.

%For an estimation of the {\em magnitude} of $\zeta$ we consider the bacteria effectively as a sphere with diameter $d=3*10^{-6}$m.
%The swimming speed is of the order $20*10^{-6}$m/s and that the solvent viscosity (e.g., water) is $10^{-3}$Nsm$^{-2}$. The resulting
%Stokes force is about 1 pN. With the maximum packing fraction in a two-dimensional (hard-disk) solution in its liquid state (i.e., below the freezing transition) $\varphi=\pi/4 (N/A)d^2\approx 0.7$, we estimate the order of the maximum $\zeta$ as $10^{-1}$.

Finally,  our analysis neglects any fluctuations of the concentration of active particles in the suspension.
Clearly there are conditions, where the dynamics of the concentration field can strongly influences the other variables \cite{Giomi}.
The goal of the subsequent analysis is to investigate the influence of the parameters $\lambda$ and $\zeta$
on experimentally accessible rheological properties of the active system. In doing this we focus
on the passive, isotropic state rather than on the metastable regime at the isotropic-nematic transition which was studied numerically in  Ref. \cite{YeomansCates}.
%Our approach focus on the passive isotropic not on the metastable regime between isotropic-to nematic transition as investigated numerically in  Ref. \cite{YeomansCates}.
%{\bf Here we should also briefly comment on differences of our model equations (\ref{modelEq}) and (\ref{stressEq}) to other studies of the rheology
%(see e.g., Equation (2) in Ref. \cite{YeomansCates}).}
%
% Here we treat \lambda and \zeta as additional free parameters ???
%Microscopic considerations presented in [??] indicate order of magnitude, sign,
%relation between \lambda and \zeta ???

\section{Analytical Results}
\label{secviscosity}
The flow geometry is given by two infinitely extended plates in the $x$-$z$-plane, which are
separated by a distance $2h$ along the $y$ direction. We apply a planar  Couette flow by moving the upper plate by $+v^w$ and the lower by $-v^w$, yielding a velocity profile
$v_\mu=(v(y),0,0)$. Under these conditions the strain-rate tensor simplifies for a linear flow profile to $\Gamma_{\mu \nu}=\partial_y v_x/2 \delta_{\mu y} \delta_{\nu x}
=\dot{\gamma}/2\delta_{\mu y} \delta_{\nu x}$, where $\dot{\gamma}=\partial v/\partial y$ is the shear rate.

To rewrite the tensorial equations of motion Eqs.~(\ref{modelEq}) and (\ref{stressEq}) in a more explicit way,
we express the second-rank tensors $Q_{\mu \nu}$ and $\sigma_{\mu \nu}$, which have five independent components,
in a tensor  basis, e.g., $Q_{\mu \nu}=\sum_{a=0}^4 Q_\mathrm{a}T_{\mu \nu}^\mathrm{a}$ \cite{Kaiser}. The basis tensors are defined by
\begin{eqnarray}
\label{basis}
T^0_{\mu \nu}&:=& \sqrt{\frac{3}{2}} \stl{e_\mu^z e_\nu^z}, \, \,\, T^1_{\mu \nu}:= \sqrt{\frac{1}{2}} (e_\mu^x e_\nu^x-e_\mu^y e_\nu^y), \, \,\,T^2_{\mu \nu}:=\sqrt{2} \stl{e_\mu^x e_\nu^y}  \nonumber \\
T^3_{\mu \nu}&:=&\sqrt{2} \stl{e_\mu^x e_\nu^z} ,\, \,\, T^4_{\mu \nu}:= \sqrt{2} \stl{e_\mu^y e_\nu^z},
\end{eqnarray}
where, e.g., $e_\mu^z$ is the $\mu$-component of the unit vector in $z$-direction.

In the subsequent analysis we assume that the passive system is isotropic in the absence of external flow (i.e. $Q_{\mu \nu}=0$). In this case the order-parameter expansion in the equilibrium free energy can be truncated after the first (quadratic) term, yielding
only the linear contribution (i.e., $\Phi_{\mu \nu}\approx A Q_{\mu \nu}$) in the dynamic equation for $Q_{\mu \nu}$. We also assume, as a first approximation, that the system
is spatially homogeneous, that is, $\triangle Q=0$. The resulting, homogeneous
equation of Eq.(\ref{modelEq} ) for the alignment tensor $Q_{\mu \nu}$ decouples into equations for the symmetry-adapted (a$=0,1,2$) and symmetry-breaking (a$=3,4$) components. The symmetry-breaking components give no contributions to our rheological quantities.  Therefore, we focus on the symmetry-adapted components, that is,
\begin{equation}
\label{AnsatzQ}
Q_{\mu \nu}=Q_0 T^0_{\mu \nu}+Q_1 T^1_{\mu \nu}+Q_2 T^2_{\mu \nu}.
\end{equation}
A similarly decoupling occurs for the stress tensor, yielding
\begin{equation}
\sigma_{\mu \nu}=\sigma_0 T^0_{\mu \nu}+\sigma_1 T^1_{\mu \nu}+\sigma_2 T^2_{\mu \nu}.
\end{equation}
Inserting Eq.~(\ref{AnsatzQ}) into Eq.~(\ref{modelEq}) we obtain the following set of equations describing the orientational dynamics in a homogeneous, isotropic active system:
\begin{eqnarray}
\label{ordersystem}
\Gamma Q_1+\frac{1}{\sqrt{3}}\kappa \Gamma  Q_0 +Q_2 & = &
- \frac{\tau_\mathrm{Qp}}{\tau_\mathrm{Q}}\Gamma + \lambda \frac{\tau_\mathrm{Q}}{A} Q_2 \nonumber \\
Q_1 &=&\Gamma  \, \, Q_2 + \lambda \frac{\tau_\mathrm{Q}}{A} Q_1 \nonumber\\
Q_0& =&-\frac{1}{\sqrt{3}}\kappa \Gamma \, \, Q_2+ \lambda \frac{\tau_\mathrm{Q}}{A} Q_0,
\end{eqnarray}
where $\Gamma=\tau_\mathrm{Q}\dot{\gamma}/A$.
Equations~(\ref{ordersystem}) can be solved for the three symmetry-adapted components yielding, e.g.,
\begin{eqnarray}
\label{Q2}
 Q_2(\Gamma) =
-\frac{\tau_\mathrm{Qp}}{\tau_\mathrm{Q}}\frac{\Gamma}{1-\lambda^*+(1-\frac{1}{3}\kappa^2)\Gamma^2\left(1-\lambda^*\right)^{-1}}.
\end{eqnarray}
Similar expressions result for $Q_0$ and $Q_1$. In Eq.~(\ref{Q2}) we have introduced the dimensionless active parameter $\lambda^*=\lambda \tau_\mathrm{Q}/A$ related
to the aligning term in Eq.~(\ref{modelEq}).
Interestingly, there is no dependence of $Q_2$ on the other parameter related to activity, $\zeta^*= \zeta/A$, which appears in the pressure equation. We note that this is  not a consequence of the linearization, but rather of our assumption of spatial homogeneity (no backflow).

For liquid crystal polymers the shape parameter $\kappa$ depends on the coupling strength between orientation and fluids strain rate. In the kinetic approach $\kappa$ can be related to the relaxation times $\tau_\mathrm{Qp}$ and $\tau_\mathrm{Q}$ \cite{HessPhysica}, i.e.  $\kappa  = -\sqrt{15} \tau_\mathrm{Qp}/(7 \tau_\mathrm{Q})$. Note, in the description of passive liquid crystal polymers the entropy production has to be positive and Onsager's symmetry relations apply. As a result $\tau_\mathrm{p}, \tau_\mathrm{Q}>0$, but $\tau_\mathrm{Qp}$ can have either sign depending on the particles shape. For rod-like particle suspensions the relaxation time $\tau_\mathrm{Qp}$ is negative and for disk-like particles is positive, respectively. For active suspensions Onsager's symmetry relations may be broken such that $\tau_\mathrm{Q}<0$ is possible. However, for simplicity we only discuss $\tau_\mathrm{Q} > 0$.

Some other features of Eq.~(\ref{Q2}) can be seen immediately.
In the limit $\Gamma\rightarrow 0$ (no external shear flow)
$Q_2$ vanishes (as do the other components),
consistent with our assumption of a passive isotropic equilibrium system. A special situation occurs for
$\lambda^*\rightarrow 1$, which corresponds to
approaching the stability limit of the isotropic phase [see Eq.~(\ref{linear_term})]. At the same time, the order parameter as determined from Eq.~ (\ref{modelEq}) increases to infinity. This behavior contradicts the physical picture according to which the order parameter is restricted to a finite (saturation) value related to perfect alignment. Therefore, the limit $\lambda^{*}\rightarrow 1$ corresponds to a situation where
the truncation of the LG free energy after the quadratic term cannot justified any more, i.e., higher order terms need to be included to guarantee the existence of a minimum. Later, we will see that this divergence gives rise to a divergence of the viscosity for both extensile and contractile suspensions. A divergence for contractile suspensions was predicted by Marchetti and Liverpool \cite{Liverpool, Marchetti4, Marchetti5,Marchetti6} and confirmed  in numerical simulations by Cates et al. \cite{YeomansCates}.

For extensile suspensions the viscosity drops to zero \cite{YeomansCates}. As we will later see, a different behavior is predicted by our model where the viscosity diverges to minus infinity. This unphysical result effectively restricts our study to values of $\lambda < 1$.

\begin{table}[ht]
\caption{Parameters} % title of Table
\centering % used for centering table
\begin{tabular}{|c c c c|} % centered columns (4 columns)
%heading
\hline % inserts single horizontal line
$\tau_\mathrm{Qp}$ & $\tau_\mathrm{p}$ &  $\tau_\mathrm{Q}$ & $\kappa$     \\
\hline % inserts single horizontal line
         -0.1      &  0.1              &  0.2               & 0.37
           \\ [1ex] % [1ex] adds vertical space
\hline %inserts single line
\end{tabular}
\label{table:para} % is used to refer this table in the text
\end{table}
%{\bf For the discussion of the viscosity: Maybe we should show ''constitutive relations'' (such as in Fig. 1 in Ref.~\ref{YeomansCates}). Also, do we see (as they do) a divergence
%of the viscosity at the I-N spinodal for contractile systems?)}\\
%
%{\bf From here on I did not change the text}.\\
%\begin{figure}
%\begin{center}
%\begin{tabular}{c}
%\psfig{figure=Q_Wact_homo_1.eps,width=4.4in}
%\end{tabular}
%\end{center}
%\caption{The order parameter components vs. shear rate $\Gamma$ for the parameter values $ \tau_\mathrm{Qp}=-0.1, \tau_\mathrm{p}=0.1,\tau_\mathrm{Q}=0.2, \lambda = 0., \kappa=0.4$.}
%\label{act1.fig}
%\end{figure}

\begin{figure}
\begin{center}
\begin{tabular}{c c}
\psfig{figure=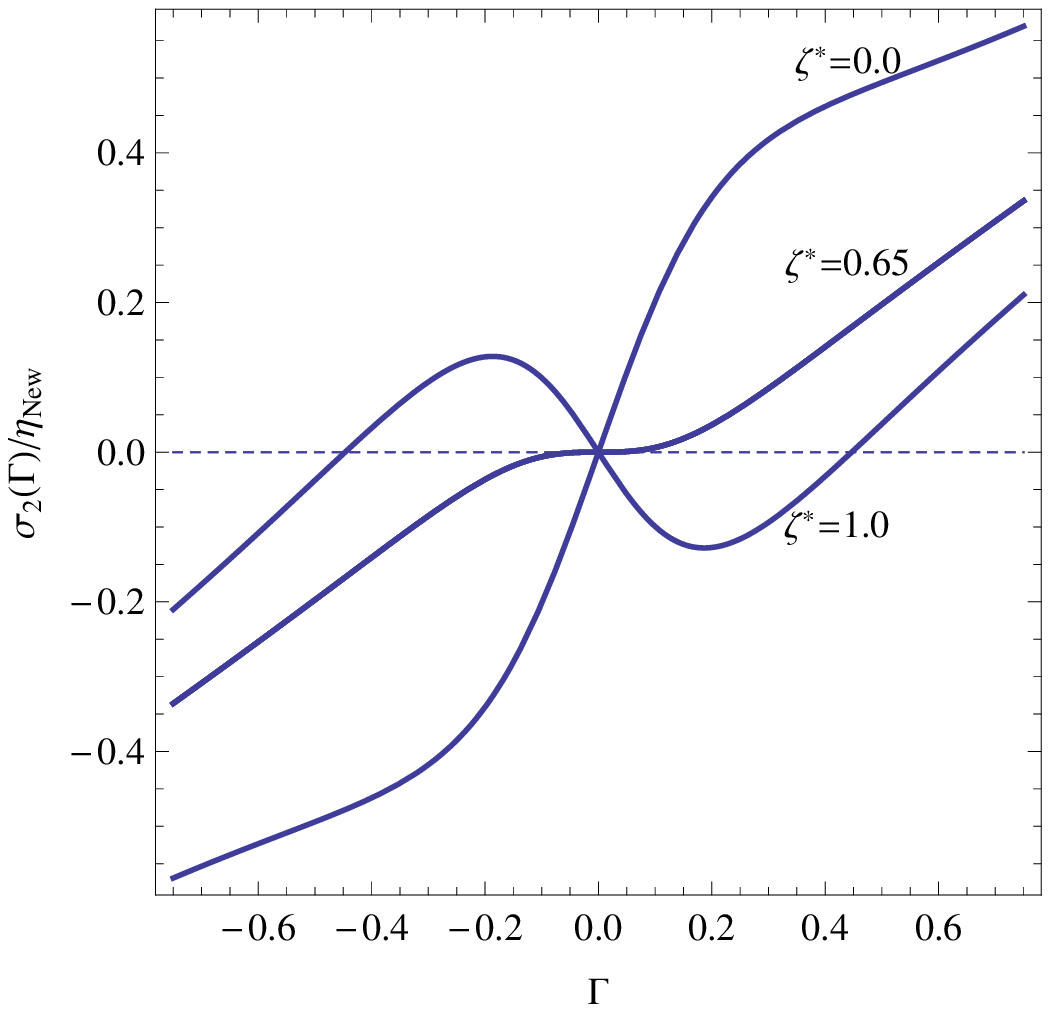,width=2.3in} &
\psfig{figure=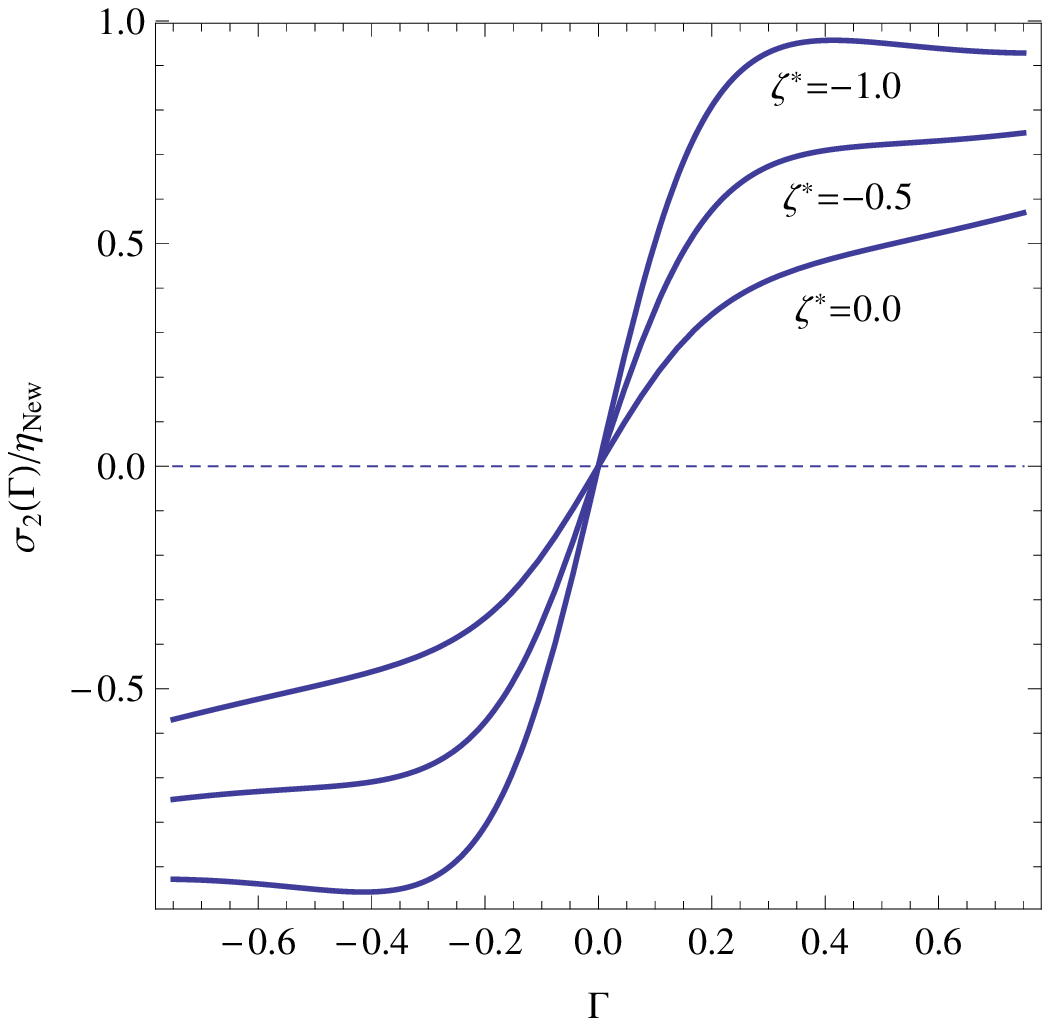,width=2.3in}\\
\end{tabular}
\end{center}
\caption{(Color online) The shear stress vs. shear rate $\Gamma$ for $\lambda^*=0.7$. The remaining parameters are given in table \ref{table:para}. Left: Extensile suspensions show a plateau and negative stress close to zero shear rate. Right: Contractile suspensions are characterized by a yield stress.}
\label{act1.fig}
\end{figure}

\subsection{Apparent shear viscosity}
The rheological properties of non-Newtonian fluids can be captured experimentally by measuring the shear viscosity and the normal stress differences.
In the geometry used here, the shear viscosity is determined by the $xy-$ component of the stress tensor, i.e. $\stl{\sigma_\mathrm{xy}}=\sqrt{2}\sigma_2$. In terms of the shear rate $\Gamma$ the shear stress reads
\begin{eqnarray}
\sigma_2&=&\eta_\mathrm{iso} \dot{\gamma} -\frac{\rho}{m}k_\mathrm{B}T A\left(\frac{\tau_\mathrm{Qp}}{\tau_\mathrm{Q}}(1+\frac{\tau_\mathrm{Q}}{\tau_\mathrm{Qp}}\zeta^*)Q_2\right.
 -\left. \frac{2}{3}\kappa^2 \frac{\Gamma}{1-\lambda^*} Q_2^2 \right),
\end{eqnarray}
where $\eta_\mathrm{iso}=(\rho/m)k_\mathrm{B} T \tau_\mathrm{p}\left(1-
\tau_\mathrm{Qp}^2/\left(\tau_\mathrm{Q}\tau_\mathrm{p}\right)\right)$ is
the Newtonian viscosity in a system with $Q=0$ (i.e., in an isotropic state or for systems of spherical
particles).

The apparent shear viscosity $\eta$ is determined by the ratio of shear stress versus the shear rate. For our analysis we scale $\eta$ with the (passive) first Newtonian viscosity $\eta_\mathrm{New}=\rho/m k_\mathrm{B} T \tau_\mathrm{p}$, yielding
%Otherwise, we scale to the zero shear rate $\eta_0=\eta(\dot{\gamma}=0)$ to focus on the relation between the orientation an the viscosity.
 %The non-Newtonian viscosity $\eta$ scaled with respect to $\eta_\mathrm{New}$ yields
\begin{equation}
\eta^*=\frac{\eta}{\eta_\mathrm{New}}=1+ \frac{\tau_\mathrm{Qp}^2}{\tau_\mathrm{Q} \tau_\mathrm{p}}\left(\frac{1-\lambda^*+(1+\frac{1}{3}\kappa^2)\frac{\Gamma^2}{1-\lambda^*}}{\left[1-\lambda^*+(1-\frac{1}{3}\kappa^2)\frac{\Gamma^2}{1-\lambda^*}\right]^2 }+\frac{\tau_\mathrm{Q}}{\tau_\mathrm{Qp}}\frac{\zeta^*}{1-\lambda^*+(1-\frac{1}{3}\kappa^2)\frac{\Gamma^2}{1-\lambda^*}}-1\right).
 \end{equation}
 For vanishing activity parameters $\lambda^*=0$ and $\zeta^*=0$ the apparent shear viscosity of a passive liquid crystal is recovered \cite{Borgmeyer_Hess}.

\begin{figure}
\begin{center}
\begin{tabular}{c}
\psfig{figure=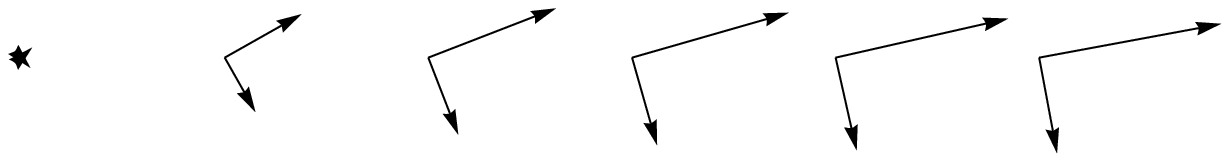,width=3.8in}\\
\psfig{figure=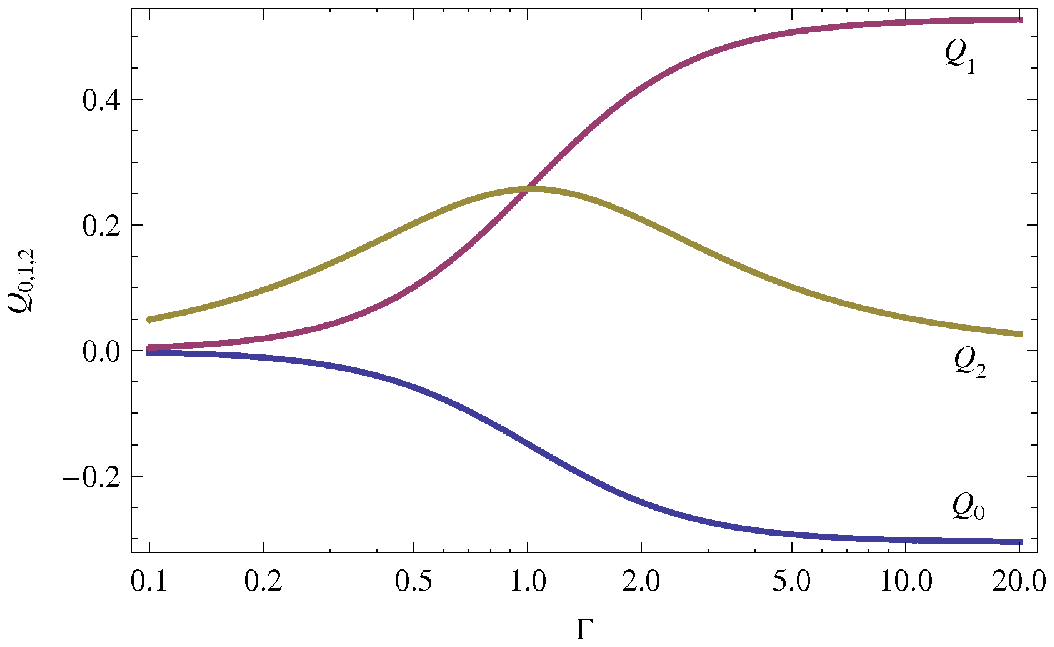,width=4.0in}\\
\psfig{figure=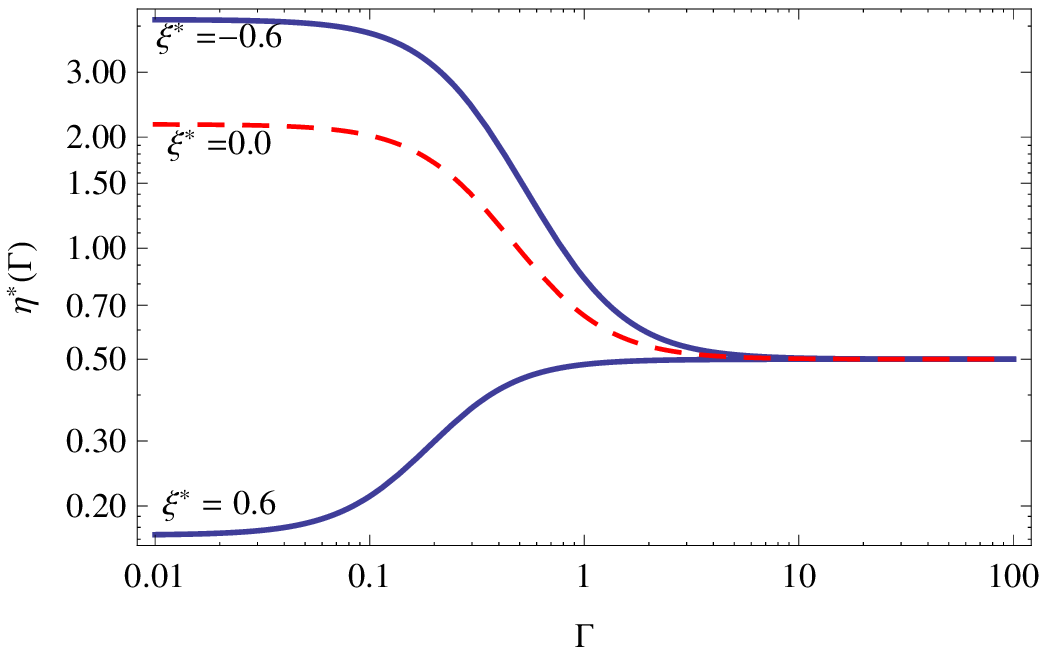,width=4.0in}\\
\end{tabular}
\end{center}
\caption{(Color online)
Top: The components of the order parameter $Q_i$ for parameters given in \ref{table:para}. The arrows above the figure show the eigenvectors of ${\bf Q}$. The length is given by the magnitude of the eigenvalues.\\
Bottom: The scaled viscosity $\eta^*$ vs. shear rate $\Gamma$ for $\zeta^*=-0.6, 0.0, 0.6$ and $ \lambda^*=0.7$. The remaining parameters given in table \ref{table:para}. The red dashed line indicates shear thinning for the passive suspension.}
\label{act12.fig}
\end{figure}
%
%\begin{figure}
%\begin{center}
%\begin{tabular}{c}
%%\psfig{figure=H_Wact_homo_1.eps,width=4.6in}\\
%\psfig{figure=H_lamb_homo.eps,width=4.0in}\\
%\psfig{figure=H_lamb1_homo.eps,width=4.0in}
%\end{tabular}
%\end{center}
%\caption{The scaled viscosity $\eta^*$ vs. shear rate $\Gamma$ for the parameter values $ \tau_\mathrm{Qp}=-0.1, \tau_\mathrm{p}=0.1,\tau_\mathrm{Q}=0.2, \kappa=1.3$, $\lambda=0$ and (upper figure) $\zeta=-2.5, -0.5,0, 0.5$.
%In the lower panel the active orientational parameter is $\lambda=-0.5,0,0.5$ varied for fix $\zeta=0$. The red dashed line of the upper and lower panel give the graph for $\zeta=0$ and $\lambda=0$, respectively}
%\label{act3.fig}
%\end{figure}
In Fig. \ref{act1.fig}, the flow curve of our model is shown for extensile and contractile suspensions.
For extensile suspensions there exists a threshold $\zeta_c^*=0.55$ above which the shear stress shows a plateau. A plateau in the constitutive curve is frequently used to explain shear banding flow instabilities in passive complex fluids that have been studied for a long time \cite{Fielding}. The difference here is that the plateau occurs around the zero shear point and the related instability might generate shear bands with velocities directed in the opposite direction. At the same time the viscosity vanishes reminiscent to a {\em superfluidic} state as discussed in \cite{YeomansCates, Giomi1}.

In a second possible scenario, at a critical nonzero stress $-\sigma_0$ ($\sigma_0$) the shear rate jumps from a positive (negative) to a negative (positive) value  forming a hysteresis \cite{Giomi1}. Hysteresis effects around a positive shear rate value can be observed for anisotropic complex fluids \cite{Klapp_hyst}. In contrast to the passive system for active suspensions the hysteresis {\em enclose} the zero shear rate point. This means that in a shear stress controlled experiment starting with $\sigma>\sigma_0$ there is a threshold before the direction of the velocity can be flipped. A similar effect is not observed in passive suspensions. The threshold is related to the extra active stress due the active particles orientation and interaction to the solvent. Contractile suspensions do not show a plateau, but a yield stress which becomes more and more pronounced for higher values of $\zeta^*$.

Passive liquid-crystalline systems typically display shear thinning, that is, the shear viscosity decreases with increasing $\Gamma$. This is indicated
by the red dashed line in Fig.~\ref{act12.fig}. The shear thinning results from the coupling of the flow gradient and the orientational degrees of freedom, yielding
flow-alignment of the suspension. In our model all three components of the tensorial order parameter are effected. The eigenvectors and eigenvalues indicates the biaxiality of the flow aligned state. The arrows above  Fig. \ref{act12.fig} display the eigenvectors multiplied by its eigenvalues. A small angle is enclosed between the principal director (the eigenvector corresponding to the highest eigenvalue) and the flow direction.
The lower part of Fig.~\ref{act12.fig} shows the viscosity for extensile and contractile suspensions. Obviously, the behavior if $\eta$ depends very strongly on $\zeta$ at low shear rates, but becomes essentially independent at high shear rates. The intermediate regime is characterized by shear thinning or -thickening behaviour depending on the actual value of $\zeta$.

For passive liquid crystal polymers in the zero shear limit the viscosity is equal to the the Newtonian viscosity $\eta_\mathrm{iso}$. However, for active suspensions the zero shear viscosity also depends on active parameters, that is
\begin{equation}
 \eta^*_0=1-\frac{\tau_\mathrm{Qp}^2
}{\tau_\mathrm{Q}\tau_\mathrm{p}} \left(\frac{\lambda^*+\frac{\tau_\mathrm{Q}}{\tau_\mathrm{Qp}}\zeta^*}{\lambda^*-1}\right).
\end{equation}
It is remarkable that in the active case the shear viscosity at $\Gamma=0$ strongly differs from the Newtonian case, indicating the non-equilibrium character of active suspensions.

For contractile suspensions the shear thinning is enhanced and diminished for extensile. For extensile suspensions there is transition to shear thickening when the zero shear viscosity becomes smaller than the second Newtonian viscosity $\eta^*_\infty=1-\tau_\mathrm{Qp}^2/(\tau_\mathrm{Q} \tau_\mathrm{p})$.  For very high extensile strength the zero shear viscosity can vanish related to the plateau of the shear stress rate curve Fig. \ref{act1.fig}. The effect of reduced viscosity was also discussed for a two-dimensional model of dilute bacterial suspensions in \cite{Heines}.
As mentioned before the high shear rate viscosity reaches the second Newtonian viscosity independent on the activity.

In our model the shape parameter $\kappa$ influences the results only marginally. In the special case $\kappa=0$
the shear viscosity simplifies to
\begin{equation}
\eta^*=\eta^*_\infty+\frac{\eta^*_0-\eta^*_\infty}{1+(\tau_r \dot{\gamma})^2},
\label{evis}
\end{equation}
%Here $\eta_1=\frac{1-\eta_0}{1-\eta_\infty}$.
where $\tau_r$ is a characteristic time defined as $\tau^{-1}_r=A/\tau_\mathrm{Q}-\lambda^*=6 D_r-\lambda^*$. Therefore, the activity parameter $\lambda^*$ modifies the self-rotational diffusion constant $D_r$.  The apparent viscosity Eq. ( \ref{evis}) can be related to a simple rheological shear thinning model, that is, the Cross model \cite{Walters}:
\begin{equation}
\eta=\eta_\infty+\frac{\eta_0-\eta_\infty}{1+(C\dot{\gamma})^m}.
\end{equation}
For our model we find $C=\tau_r$ and $m=2$.

The simplified formula (17) gives us the possibility to link experimentally accessible
quantities to our theoretical parameters $\lambda^{*}$ and $\zeta^{*}$.
Specifically, we need from experiments the zero-shear viscosity, the high-shear viscosity $\eta_\infty$, the rotational diffusion constant, and the relaxation time $\tau_r$. The latter can be determined from a fit of the experimental shear viscosity
to Eq. (16). The rotational diffusion constant $D_r$ has to be measured in an independent experiment, see e.g. \cite{Lettinga} for rod-like fd-viruses. From these
quantities, the activity parameter $\lambda$ follows via the relation
$\lambda^* = 6 D_r -\tau_r^{-1}$. Next, the active force acting on the fluid modelled by $\zeta^*$ is related to $\lambda^*$ via

\begin{equation}
 \frac{\tau_\mathrm{Q}}{\tau_\mathrm{Qp}} \zeta^*=(\eta^*_1-1)\lambda^*-\eta^*_1,
\end{equation}
where we defined the viscosity $\eta^*_1=\frac{1-\eta_0^*}{1-\eta^*_\infty}$. Here the ratio of the relaxation times can be related to the active particles shape \cite{Peterlin}, i.e. $\frac{\tau_\mathrm{Qp}}{\tau_\mathrm{Q}} = -\sqrt{\frac{3}{2}} R $.  The coefficient $R$ measures the non-sphericity of particles. For isolated ellipsoids with  the semi-axes $a=b$, $c$ and the aspect ratio $q=c/b$, one has $R=(q^2-1)/(q^2+1)$.

There is a remarkable symmetry between the particles shape and response to the flow. The rheological properties do not distinguish between extensile disk-like particles (contractile, rod-like particle) and contractile rod-like particles (extensile, rod-like particle), respectively.

\subsection{Normal stress differences}
\label{stress}
The appearance of normal stress differences indicate strong non-Newtonian behaviour and is related to surprising rheological effects, e.g., the Weissenberg effect or swelling jets
\cite{Walters}. In our theoretical description we can compute the normal stress differences from the components
of the stress tensor via $N_1=\sigma_{xx}-\sigma_{yy}$ and  $N_2=\sigma_{yy}-\sigma_{zz}$. In our notation
\begin{equation}
 N_1= 2 \sigma_1, \, \, \, \, \, \, N_2=-\sqrt{3}\sigma_0 - \sigma_1.
\end{equation}
The analytical expression of the stress tensor components are given by
\begin{eqnarray}
\sigma_1 &=&-\frac{\rho}{m}k_\mathrm{B}T A\left(\frac{\tau_\mathrm{Qp}}{\tau_\mathrm{Q}}(1+\frac{\tau_\mathrm{Q}}{\tau_\mathrm{Qp}}\zeta^*)\frac{\Gamma Q_2}{1-\lambda^*}\right. -\left. \frac{2}{3}\kappa^2  \frac{\Gamma^2}{(1-\lambda^*)^2} Q_2^2  \right) \\
\sigma_0&=&-\frac{\rho}{m}k_\mathrm{B}T A\left(-\frac{\kappa}{\sqrt{3}}\frac{\tau_\mathrm{Qp}}{\tau_\mathrm{Q}}(1+\frac{\tau_\mathrm{Q}}{\tau_\mathrm{Qp}}\zeta^*)\frac{\Gamma}{1-\lambda^*} Q_2\right. \nonumber \\
 &+&\left. \frac{1}{\sqrt{3}}\kappa \left( Q_2^2+(1-\frac{1}{3}\kappa^2) \frac{\Gamma^2}{(1-\lambda^*)^2} Q_2^2\right) \right)
\end{eqnarray}
From these relations it follows that
\begin{equation}
\frac{N_1}{\eta_\mathrm{New}}= 2  \frac{\tau_\mathrm{Qp}^2 A}{\tau_\mathrm{Q}^2 \tau_\mathrm{p}}\left(\frac{1-\lambda^*+(1+\frac{1}{3}\kappa^2)\frac{\Gamma^2}{1-\lambda^*}}{\left[1-\lambda^*+(1-\frac{1}{3}\kappa^2)\frac{\Gamma^2}{1-\lambda^*}\right]^2 }+\frac{\tau_\mathrm{Q}}{\tau_\mathrm{Qp}}\frac{\zeta^*}{1-\lambda^*+(1-\frac{1}{3}\kappa^2)\frac{\Gamma^2}{1-\lambda^*}}\right)\frac{\Gamma^2}{1-\lambda^*}.
\end{equation}
Similarly $N_2$ is given by
\begin{eqnarray}
\frac{N_2}{\eta_\mathrm{New}}&=&
 \frac{\tau_\mathrm{Qp}^2 A}{\tau_\mathrm{Q}^2 \tau_\mathrm{p}}
 \left(\frac{2\kappa+\frac{\tau_\mathrm{Q}}{\tau_\mathrm{Qp}}\zeta^* \kappa-\frac{\tau_\mathrm{Q}}{\tau_\mathrm{Qp}} \zeta^*}{1-\lambda^*} \frac{\Gamma^2}{1-\lambda^*+(1-\frac{1}{3}\kappa^2)\frac{\Gamma^2}{1-\lambda^*}}\right.
\\
 \nonumber
&-& \left.  \frac{ 1-\lambda^*+(1+\frac{1}{3} \kappa^2) \frac{\Gamma^2}{1-\lambda^*}}{\left[1-\lambda^*+(1-\frac{1}{3}\kappa^2)\frac{\Gamma^2}{1-\lambda^*}\right]^2}\frac{\Gamma^2}{1-\lambda^*} \right). \nonumber
\end{eqnarray}
%
%
%The infinite shear limit for $\kappa=0$ reads
%\begin{equation}
% \lim_{\Gamma \rightarrow \infty}N_1=2 \frac{A \tau_\mathrm{Qp}}{\tau_Q^2\tau_p}\frac{1-\frac{\tau_\mathrm{Q}}{\tau_\mathrm{Qp}}(\lambda^*-1)\zeta^*}{1-\lambda^*}
%\end{equation}
In passive, nematic liquid-crystal polymers subject to a shear flow the first normal stress difference
displays a different sign depending on the shear rate. The resulting change of sign
was first observed in experiments by Porter and Kiss \cite{Porter_Kiss} (see also \cite{Cementwala,Tao,Mewis} for further experiments and numerical simulations).

\begin{figure}
\begin{center}
\begin{tabular}{c c}
\psfig{figure=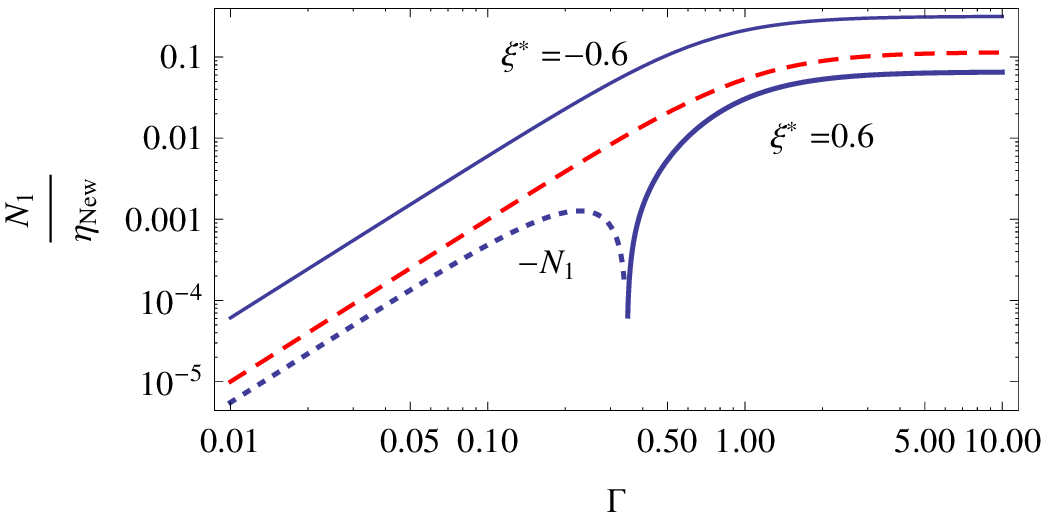,width=4.0in} \\
\psfig{figure=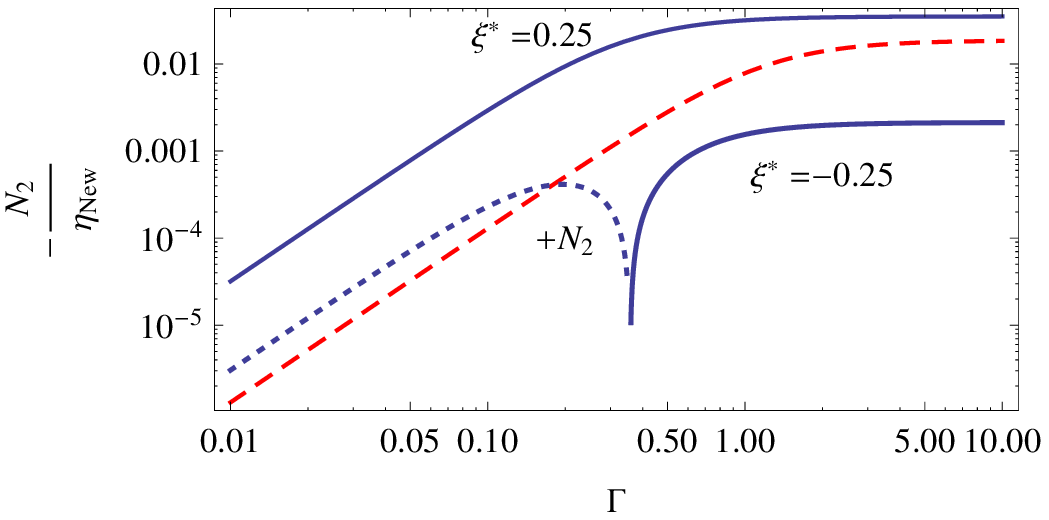,width=4.0in}
\end{tabular}
\end{center}
\caption{(Color online)
Top:
The first normal stress difference changes its sign for extensile ($\zeta^* = 0.6$) active particle suspensions. The dotted line illuminates negative values of $N_1$. The curve for contractile suspensions is shifted with respect to the curve of passive suspensions (red dashed line).
Bottom:
The sign change of the second normal stress difference appears for contractile suspensions. The dotted line illuminated the positive values of $N_2$. For extensile suspensions there is no sign change, but a shift with respect to the passive curve (red dashed line). The remaining parameters of both figures are shown in table \ref{table:para}.}
\label{N.fig}
\end{figure}

On the other hand, isotropic liquid crystal polymers do not show that sign change. The first (second) normal stress difference for isotropic liquid crystal polymers is positive (negative) and increases (decreases) with increasing shear rate.
In contrast, active suspensions deep in the isotropic phase can change the sign of the normal stress differences. In Fig. \ref{N.fig} the dashed red line shows the dependence of $N_1$ and $N_2$ on the shear rate for passive liquid-crystal polymers (i.e., $\lambda=\zeta=0$).

\begin{figure}
\begin{tabular}{c}
\psfig{figure=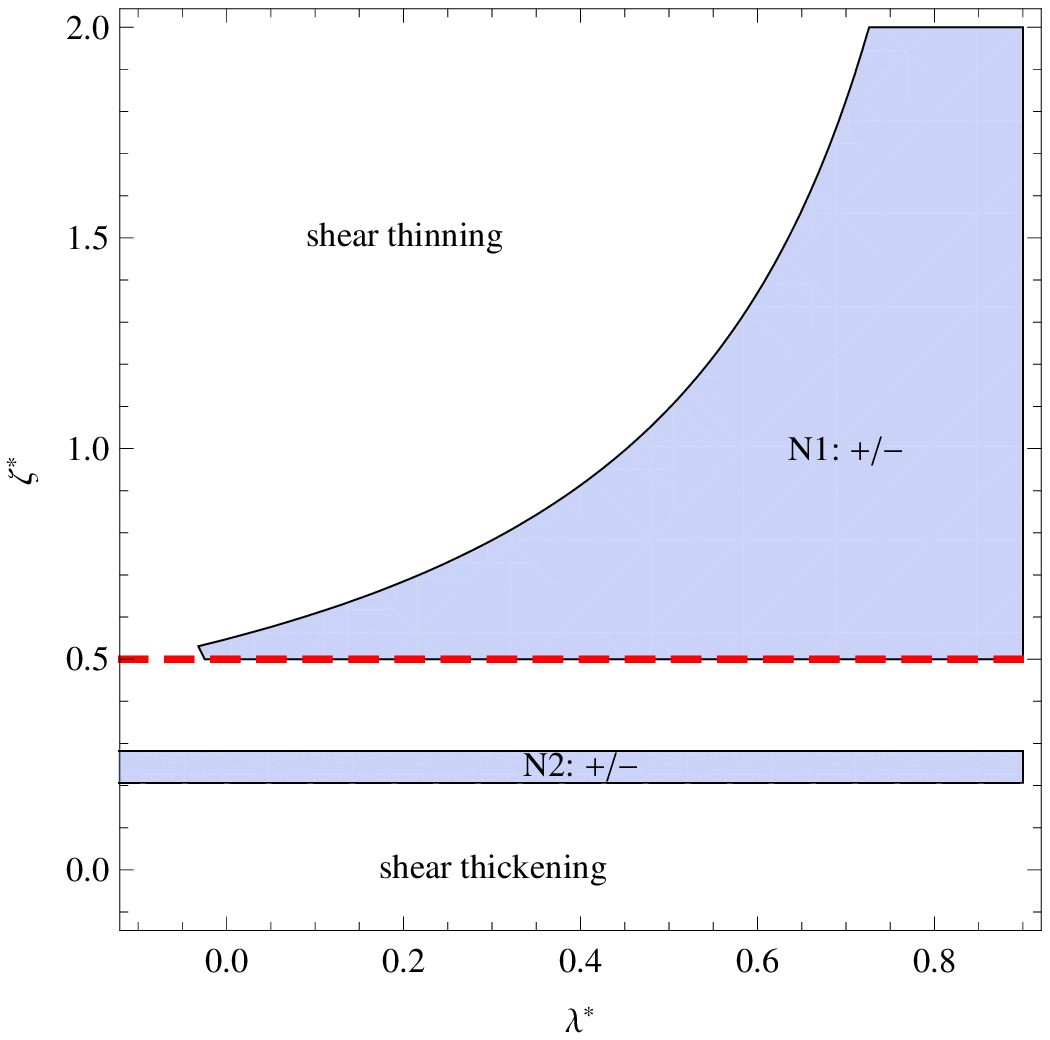,width=4in}\\
\psfig{figure=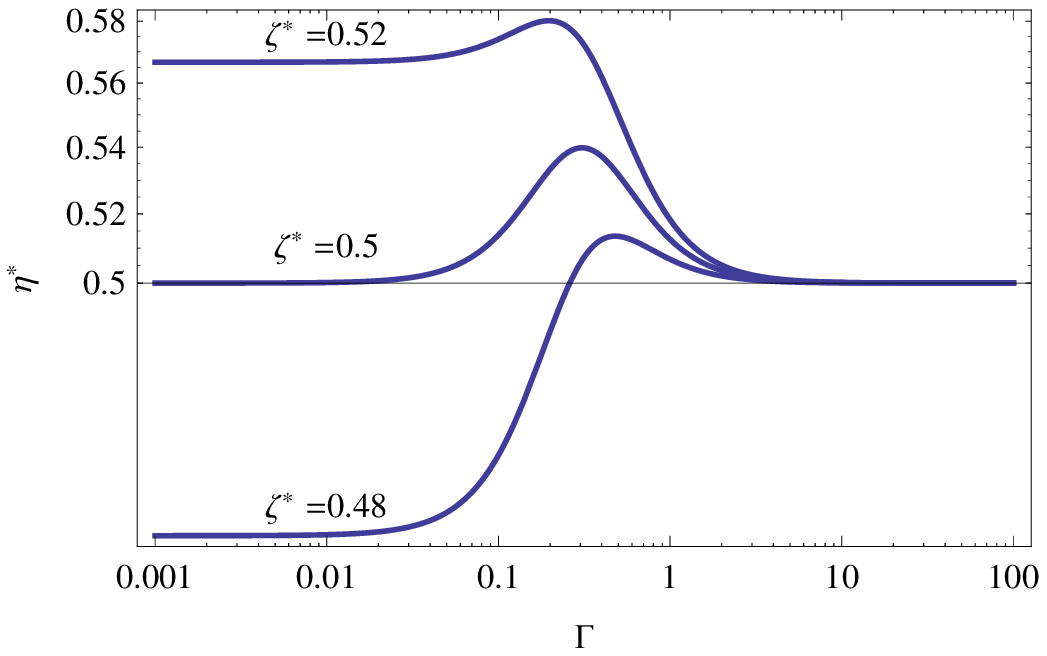,width=4in}
\end{tabular}
\caption{(Color online)Top: Rheological diagram of active suspensions in the plane spanned by the
two activity parameters. The red dashed indicated the activity parameter $\zeta_\mathrm{c}$ that divides the plane in a shear thinning and a shear thickening region. For shear thickening there is a region that shows sign change of the first normal stress difference. On the other hand, in the shear thinning regime there exist a parameter range for sign change of the second normal stress difference. The parameter set: $A=0.2$ and \ref{table:para}.\\
Bottom: The apparent shear viscosity close to $\zeta_\mathrm{c}$ shows a shear thickening and thinning behaviour. The parameter value of $\lambda=0.7$.}
\label{N1.fig}
\end{figure}

For contractile suspensions there is a parameter range where the sign of the first normal stress difference changes at low shear rates. The upper panel of Fig \ref{N1.fig} shows the relevant parameter range in the $\lambda^*-\zeta^*$-plane. The red dashed line divides the plane in two regions, one of shear thinning and one of shear thickening. The sign change of $N_1$ appears in the shear thinning regime only. The relevant region growth with increasing $\lambda^*$ consistent with the limiting case of passive nematic liquid crystal polymers ($\lambda^*->1$).

The second normals stress difference for passive suspensions is negative for all shear rates (red dashed line, Fig. \ref{N.fig}). However, for active suspension there is a small parameter regime in the shear thickening region (\ref{N1.fig}) that shows a change in sign as shown in Fig \ref{N.fig}.
Outside the relevant parameter regions the normal stress difference have the same qualitative behaviour as their passive counterparts. The difference is a shift that depends on both parameters $\lambda^*$ and $\zeta^*$. \\
The transition line from shear thickening to shear thinning in Fig. \ref{N1.fig} is not sharp. For a fixed  $\lambda^*$ value and  increasing parameter, from $\zeta^*=-0.6$, a smooth crossover from thickening to thinning in the vicinity of the dashed red line in Fig \ref{N1.fig} (upper panel) can be observed. The smooth crossover is characterized by a small overshoot that growth and disappears as illustrated in the lower panel of Fig. \ref{N1.fig}.

In the limit of zero shear, both normal stress differences are vanishing. However, the ratio $N_2/N_1 $, given by
\begin{equation}
 \lim_{\Gamma \rightarrow 0} \frac{N_2}{N_1}=-\frac{1}{2}\left( 1-\frac{2+\frac{\tau_\mathrm{Q}}{\tau_\mathrm{Qp}}\zeta^*}{1+\frac{\tau_\mathrm{Q}}{\tau_\mathrm{Qp}}\zeta^*}\kappa \right)= \lim_{\Gamma \rightarrow 0}\frac{\Psi_2}{\Psi_1},
\end{equation}
is a constant. The so-called viscometric functions are related to the normal stress differences by
\begin{equation}
\Psi_1=N_1\gamma^2, \; \; \Psi_2=N_2\gamma^2.
\end{equation}
The ratio of the viscometric coefficients for $\kappa \approx 0.4$ in the passive limit is estimated to $\Psi_2/\Psi_1 \approx -0.1$ which is a value typical for polymeric fluids \cite{Bird}. The corresponding value for active fluids can vary significantly. For example, at $\zeta^*=0.3$ one obtains $\Psi_2/\Psi_1 \approx 0.2$ and for $\zeta^*=-0.3$, $\Psi_2/\Psi_1 \approx -0.175$. This is an additional example for the different rheological behaviour of active matter.

% \begin{figure}
% \begin{center}
% \begin{tabular}{c}
% \psfig{figure=N1_Wact_hom.eps,width=4.0in} \\
% \psfig{figure=N1_lamb_hom.eps,width=4.0in} \\
% \end{tabular}
% \end{center}
% \caption{The scaled first stress difference for parameter values $\tau_\mathrm{Qp}=0.1, \tau_\mathrm{a}=0.2, \tau_\mathrm{p}=0.1$ The activity parameter are $\zeta=-1,-0,1$, $\lambda =0$ (upper) and $\zeta=0$, $\lambda =0.4,0,-0.4$ (lower)}
% \label{d_0.fig}
% \end{figure}

\section{Conclusion}
\label{conclude}
The analytical approach of our simplified model is able to describe some significant rheological features found earlier in numerical investigations of more involved models \cite{YeomansMarenduzzo,YeomansCates}.
One important feature concerns the effect of activity on the zero-shear viscosity which can
be significantly larger (contractile) or smaller (extensile) as compared to the passive case.
Therefore, and since the high-shear viscosity is essentially independent of activity,
the intermediate shear-rate regime of
active suspensions is characterized by either strong shear thinning (contractile) or thickening (extensile). Recently, strong shear thinning was experimentally observed
\cite{Peyla} in suspensions of {\em Chlamydomonas Reinhardii}.

Furthermore, we investigated the normal stress differences and found a remarkable sign change absent in the passive case. This novel effect is characterized by the activity of the suspension. We predict the sign change for the first normal stress difference for a wide parameter rage. For example {\em Chlamydomonas Reinhardii} may show the predicted effect due to strong shear thinning and flow aligning effect similar to rod-like passive suspensions. Moreover, we have shown that the corresponding ratio of the viscometric functions deviates markedly from the passive case.

In the present study we have not considered the impact of spatially inhomogeneity on the rheology of active matter, although this may clearly be relevant \cite{Porst}. Here we have rather assumed
that inhomogeneities are small and do not change our predictions qualitatively. Indeed, we have performed some preliminary test of our predictions, using an one-dimensional, small
sine-shaped inhomogeneity. The resulting equations are rather involved and thus not given here. An explicit numerical investigation is more reasonable and will be presented elsewhere.
Note, that in principle the active parameters may also spatially dependent \cite{SoftMarenduzzo} and probably not independent of each other. However, to our knowledge there is no rigorous derivation of the parameters from the micro-swimmers that includes the swimming mechanism.

In our study we assumed that the effective interaction between active particles lead to isotropic and nematic order as, for example, migrating cells \cite{Gruler}. Similar to the hydrodynamic theory of nematic liquid crystals we coupled the generalized Navier-Stokes equations to the second rank symmetric traceless order parameter ${\bf Q}$. However, in dilute suspensions when self-driven particles show swarming motion, the symmetry of the particles orientational distribution leads to a vectorial order parameter ${\bf P}$ rather than a ${\bf Q}$ tensor.  The rheology for active suspensions related to a polarization vector ${\bf P}$ are studied by Giomi et. al.  \cite{Giomi}. Similar to our findings the model predicts shear thinning, thickening and vanishing viscosity. The effect on normal stress differences is not investigated, but we expect no significantly different results. In a further extension of the theoretical description a coupled vector-tensor model used for passive polarized nano-rod suspensions may show additional rheological effects \cite{Grandner_1,Grandner_2}.

Very recently, Saintillan used a Fokker-Planck approach to model the orientational dynamics of micro-swimmers in {\em extensional} flows. He models active stress contributions by force dipoles that are generated by swimming strokes of the individual swimmers. Depending on the extensional rate the suspension shows thinning, thickening and a negative effective viscosity similar to our results. It is remarkable that, despite different equations used in the passive part of the pressure tensor and of the orientational dynamics, all models lead to qualitatively the same rheological behavior. We expect that the active stress coupling modeled by force dipoles are responsible for the effects discussed here. In that sense shear thinning, thickening, zero effective viscosity and sign change in the stress differences are general effects in active soft matter.

We finish with a remark on the relaxation times. The relaxation times of the passive suspension are restricted due to Onsager's symmetry relations, viz $\tau_\mathrm{Q}>0$, $\tau_\mathrm{p}>0$ and $\tau_\mathrm{Q}\tau_\mathrm{p}>\tau_\mathrm{Qp}$. On the other hand active particle suspensions always dissipate energy due to the process of metabolism. Therefore, a general description needs additional variables that are related to the process of metabolism and motility of the swimmers. As a consequence, the relations for the relaxation times in the passive case may be violated.

{\Large Acknowledgments}\\
We thank R. Vogel for useful discussions. Financial support within the Collaborative Research Center
"Mesoscopically structured Composites" (Sfb 448, project B6) and the project HE5995/1-1 of the Deutsche Forschungsgemeinschaft  is gratefully acknowledged.

%%%%%%%%%%%%%%%%%%%%%%%%%%%%%%%%%%%%%%%%%%%%%%%%%%%%%%%%%%%%%%%%%%%%%
%% The appropriate \bibliographystyle and \bibliography commands
%% should be placed here.
%%%%%%%%%%%%%%%%%%%%%%%%%%%%%%%%%%%%%%%%%%%%%%%%%%%%%%%%%%%%%%%%%%%%%

\end{document}

%% file: tex_options.tex
\usepackage{graphicx}
\usepackage{dcolumn}
\usepackage{epstopdf}
\usepackage{eucal}
\usepackage[dvips]{epsfig}
\usepackage{amssymb}

\usepackage{amsmath,amsthm}
\newcommand{\be}{\begin{eqnarray}}
\newcommand{\ee}{\end{eqnarray}}
\newcommand{\bse}{\begin{subequations}}
\newcommand{\ese}{\end{subequations}}

\newcommand{\bnum}{\begin{enumerate}}
\newcommand{\enum}{\end{enumerate}}

\newcommand{\bit}{\begin{itemize}}
\newcommand{\eit}{\end{itemize}}

\newcommand{\bc}{\begin{cases}}
\newcommand{\ec}{\end{cases}}

\newcommand{\bpm}{\begin{pmatrix}}
\newcommand{\epm}{\end{pmatrix}}

\newcommand{\bvm}{\begin{vmatrix}}
\newcommand{\evm}{\end{vmatrix}}

%% file: visco_active_PRE_v1.bbl
\begin{thebibliography}{srt}
\bibliographystyle{rsc}

\bibitem{Ramaswamy}
S. Ramaswamy and R. A. Simha,
 {\em Sol. State. Comm.}, 2006, {\bf 139}, 617-622.


\bibitem{Kruse}
  K. Kruse, J. F. Joanny, F. J\"ulicher, J. Prost, and K. Sekimoto,
  {\em Phys. Rev. Lett.},2004, {\bf 92}, 078101/1-4.

\bibitem{Liverpool}
  T. B. Liverpool and M. C. Marchetti,
  {\em Phys. Rev. Lett.}, 2006, {\bf 97}, 268101/1-4.

\bibitem{Ramaswamy_Science}
  V. Narayan, S. Ramaswamy, N. Menon,
  {\em Science },2007, {\bf 6}, 105.

\bibitem{Tsimring}
  A. Kudrolli, G. Lumay, D. Volfson, and L. S. Tsimring,
  {\em Phys. Rev. Lett.}, 2008, {\bf 100}, 058001/1-4.

\bibitem{Velev}
  S. T. Chang, V. N. Paunov, D. N. Petsev and O. D. Velev,
  {\em Nature Materials},2007, {\bf 6}, 235-240.

\bibitem{Winet}
  C. Brennen and H. Winet,
  {\em Ann.  Rev. of Fluid Mech.}, 1977, {\bf 9}, 339-98.

\bibitem{Hatwalne}
   Y. Hatwalne, S. Ramaswamy, M. Rao, and R. A. Simha,
  {\em Phys. Rev. Lett.},2004, {\bf 92}, 118101/1-4.

\bibitem{Gruler}
  H. Gruler, M. Schienbein, K. Franke, and A. DeBoisfleury-Chevance,
  {\em Mol. Cryst. Liq. Cryst.},1995, {\bf 260}, 565-574.

\bibitem{Simha}
  R. A. Simha and S. Ramaswamy,
  {\em Phys. Rev. Lett.},2002, {\bf 89}, 058101/1-4.

 \bibitem{Porst}
  R. Voituriez, J. F. Joanny, and J. Porst,
  {\em Eur. Phys. Lett.},2005, {\bf 70}, 404-410.

  \bibitem{Joanny}
  R. Voituriez, J. F. Joanny, and J. Porst,
  {\em Phys. Rev. Lett},2006, {\bf 96}, 028102/1-4.

\bibitem{Rao}
  S. Ramaswamy and M. Rao,
  {\em New J. Phys.},2007, {\bf 9}, 423/1-9.

\bibitem{YeomansMarenduzzo}
  D. Marenduzzo, E. Orlandini, and J. M. Yeomans,
  {\em Phys. Rev. Lett.},2007, {\bf 98}, 118102/1-4.

\bibitem{Yeomans1}
  S. A. Edwards and J. M. Yeomans,
  {\em Eur. Phys. Lett.},2009, {\bf 85}, 18008/1-6.

\bibitem{Giomi}
L.~Giomi, M.~C.~Marchetti, and T.~B.~Liverpool,
 {\em Phys. Rev. Lett.},2008, {\bf 101}, 198101/1-4.

\bibitem{YeomansCates}
  M. E. Cates, S. M. Fielding, D. Marenduzzo, E. Orlandini, and J. M. Yeomans,
  {\em Phys. Rev. Lett.},2008,{\bf 101}, 068102/1-4.

  \bibitem{Giomi1}
 L. Giomi, T. B. Liverpool and M. C. Marchetti,
{\em  Phys. Rev. E}, 2010, {\bf 81}, 051908/1-9.

\bibitem{Fielding}
S. M. Fielding and P. D: Olmsted
{\em  Phys. Rev. Lett.}, 2003, {\bf 90}, 224501/1-4.

\bibitem{Klapp_hyst}
S. H. L. Klapp and S. Hess
{\em  Phys. Rev. E}, 2010, {\bf 81},051711/1-9.

\bibitem{Marchetti1}
  A. Baskaran and M. C. Marchetti,
  {\em Phys. Rev. E}, 2008, {\bf 77}, 011920/1-9.

\bibitem{Marchetti2}
  A. Baskaran and M. C. Marchetti,
  {\em Phys. Rev. Lett.},2008, {\bf 101}, 268101/1-4.

\bibitem{Borgmeyer_Hess}
  C. P. Borgmeyer, S. Hess,
  {\em J. Non-Equlib. Thermodyn.},1995, {\bf 20}, 259-384.

\bibitem{Heidenreich_diss}
  S. Heidenreich,
  {\em PhD-thesis, Technische Universit\"at Berlin }{}{2008}{}.

\bibitem{Heidenreich_dip}
  S. Heidenreich, S. Hess and S. H. L. Klapp,
  {\em Phys. Rev. Lett.},2009, {\bf 102}, 028301/1-4.


%%%%%%%%%%12%%%%%%%%%
\bibitem{Shelley}
  D. Saintillan and M. J. Shelley,
  {\em Phys. Rev. Lett.},2008, {\bf 100}, 178103/1-4.

\bibitem{Tsimring1}
  I. S. Aranson and L. S. Tsimring,
  {\em Phys. Rev. E},2005, {\bf 71}, 050901(R)/1-4.

\bibitem{Zimmermann}
  R. Peter, Y. Schaller, F. Ziebert, and W. Zimmermann,
  {\em New. J. Phys.},2008, {\bf 10}, 034002/1-16.

\bibitem{Chate}
  J. M. Belmonte, G. L. Thomas, L.G. Brunnet, R. M. C. De Almeida and H. Chate´,
  {\em Phys. Rev. Lett.},2008, {\bf 100}, 248702/1-4.

\bibitem{Bachand}
Hess and G.D. Bachand,
  {\em Nanotoday},2005, {\bf 8}, 22-29.

\bibitem{Marchetti3}
 A. Ahmadi, T. B. Liverpool and M. C. Marchetti,
{\em Phys. Rev. E}, 2005, {\bf 72}(R), 060901/1-4.

\bibitem{Marchetti4}
T. B. Liverpool and M. C. Marchetti,
{\em Phys. Rev. Lett.}, 2003, {\bf 90}, 138102/1-4.

\bibitem{Marchetti5}
T. B. Liverpool and M. C. Marchetti,
{\em Eur. Phys. Lett.}, 2005, {\bf 69}, 846-852.

\bibitem{Marchetti6}
T. B. Liverpool and M. C. Marchetti,
{\em Phys. Rev. Lett.}, 2006, {\bf 97}, 268101/1-4.

\bibitem{Marchetti6}
T. B. Liverpool and M. C. Marchetti,
{\em Phys. Rev. E}, 2006, {\bf 74}, 061913/1-23.


\bibitem{HessPhysica}
S. Hess,
{\em Physica A}, 1983, {\bf 118}, 79-104.


\bibitem{Kaiser}
 P. Kaiser, W. Wiese and S. Hess,
{\em J. Non-Equilib. Thermodyn.}, 1992, {\bf 17},153-169.

\bibitem{Heines}
B. M. Heines, I. S. Aranson, L. Berlyand and D. A. Karpeev,
{\em Phys. Biology}, 2008, {\bf 5},046003/1-9.

\bibitem{Porter_Kiss}
G. Porter and R. S. Kiss,
{\em J. Polym. Sci.: Polym. Symp.},1978, {\bf 65}, 193-211.

\bibitem{Cementwala}
S. Baek, J. J. Magda, and S. Cementwala,
{\em J. Rheol.},1993, {\bf 37}, 935-945.

\bibitem{Ding_Yang}
J. Ding and Y. Yang,
{\em Rheol. Acta}, 1994, {\bf 33}, 405-418.


\bibitem{Tao}
Y.-G. Tao, W. K. den Otter, J. K. G. Dhont and W.J. Briels,
{\em J. Chem. Phys.}, 2006, {\bf 124}, 134906.

\bibitem{Mewis}
P. Moldenaers and J. Mewis,
{\em J. Rheol.}, 1986, {\bf 30}, 567-584.

\bibitem{Lettinga}
P. M. Lettinga, Z. Dogic, H. Wang and J. Vermant,
{\em Langmuir}, 2005, {\bf 21}, 8048-8057.

\bibitem{Peterlin}
A. Peterlin and H. A. Stuard,
 {\em Hand- und Jahresbuch der Chemischen Physik, volume 8,p.113},
 Eds. Eucken-Wolf, 1943.

\bibitem{Walters}
H. A. Barnes, J. F. Hutton and K. Walters,
{\em An Introduction to Rheology},
Elsevier, London 1989.

\bibitem{Bird}
R. B. Bird, R. C. Armstrong, O. Hassager amd C. F. Curtiss,
{\em Dynamics of polymeric liquids, vol 1/2},
J. Wiley, New York 1977.

\bibitem{SoftMarenduzzo}
D. Marenduzzo and E. Orlandini,
{\em Soft Matter}, 2010, {\bf 6}, 774-778.

\bibitem{Peyla}
S. Rafai, L. Jibuti and P. Peyla,
{\em Phys. Rev. Lett.}, 2010, {\bf 104}, 098102/1-4.

\bibitem{Saintillan}
D. Saintillan,
{\em Phys. Rev. E}, 2010, {\bf  81}, 056307/1-10.

\bibitem{Grandner_1}
S. Grandner, S. Heidenreich, S. Hess and S. H. L. Klapp,
{\em EPJE}, 2007, {\bf  24}, 353-365.

\bibitem{Grandner_2}
S. Grandner, S. Heidenreich, P. Ilg, S. H. L. Klapp and S. Hess
{\em Phys. Rev. E}, 2010, {\bf  75}(R), 040701/1-4.

\end{thebibliography}
